\documentstyle[aps,floats,epsf,psfig,twocolumn]{revtex}
\begin{document}
\tighten
\draft 
\title{A Consistent Picture of Electronic Raman Scattering 
and Infrared Conductivity in the Cuprates}
\author{T.P. Devereaux$^{1}$\cite{byline} and A.P. Kampf$^{2}$}
\address{$^{1}$
Department of Physics, The George Washington University, Washington, DC 20052}
\address{$^{2}$
Theoretische Physik III, Elektronische Korrelationen und Magnetismus, \\
Institut f\"ur Physik, Universit\"at Augsburg, 86135 Augsburg, Germany}
\address{~
\parbox{14cm}{\rm 
\medskip
Calculations are presented for electronic Raman scattering and
infrared conductivity in 
a $d_{x^{2}-y^{2}}$ superconductor including the effects of elastic
scattering via anisotropic 
impurities and inelastic spin-fluctuation scattering. A consistent
description of experiments on optimally doped Bi-2212 is made possible
by considering the effects of correlations on both inelastic and elastic
scattering.
}}
\maketitle

\narrowtext

Impurity effects in the cuprates have played a major
role in clarifying the nature of unconventional superconductivity.
Via the change in the density of states (DOS) 
near the Fermi level, crossover behavior 
of low temperature 
transport and thermodynamics quantities could be explored as a testing
ground for unconventional pairing\cite{review}. 
Examples include the $T$ to $T^{2}$
crossover in the low temperature magnetic penetration depth $\lambda(T)$, 
$T^{3}$
to $T$ crossover in the NMR rate, and the $\omega^{3}$ to $\omega$
crossover in the low frequency $B_{1g}$ Raman response.
Simple theories based on nearly-resonant scattering in the
impurity $T-$ matrix approximation (``dirty'' d-wave)
have accounted for these behaviors seen
in materials of various quality and impurity dopings.

However, several inconsistencies arise when attempting to put together a
complete picture. First, the magnitude of the impurity scattering needed
to fit $\lambda(T)$ and the extrapolated $T=0$ resistivity
are generally much smaller than those needed to
fit the measured frequency dependence of the
infrared conductivity (IR) and the Raman response.
In particular, the impurity scattering rate $1/\tau_{imp}=2\Gamma$
needed to fit the IR on
high quality Y-123 was 100 times larger than that needed to fit microwave
measurements\cite{Quinlan96}, while
values of $\Gamma/\Delta_{0}$ ranging up to 0.5 were
needed to interpret the Raman data taken on Bi-2201\cite{bi2201}, and some
samples of Tl-2201\cite{tl2201} and Hg-1223\cite{hg1223}.
Second, the ``universal'' dc conductivity
$\sigma(\omega\rightarrow 0,T=0)=(e^{2}/2\pi\hbar) \xi/a$\cite{pal} has not
been convincingly observed and the low temperature variation of
the conductivity $\delta\sigma(T)$ changes slower than $T^{2}$ 
\cite{newhardy}. Third, the transition temperature
T$_{c}$ is only moderately reduced by planar impurities compared
to Abrikosov-Gorkov theory\cite{Radtke} (by about a factor of
two to three). Lastly, the residual resistivity obtained for Zn doped
Y-123 suggests that a strong contribution is present in $d-$wave
scattering channels\cite{Walker}.

These inconsistencies reveal that the usual treatment of point-like $s-$wave
impurities in a $T-$matrix approach may be too naive and neglects
electronic correlations.
Experimental evidence from transport measurements in Zn doped YBCO
have revealed that the a single impurity embedded in the CuO$_{2}$ plane
disturbs the local environment and
yields an effective scattering cross section diameter
of Zn$^{2+}$ of 4.2$\AA$\cite{Chien} and is thus extended. 
Theoretical studies have shown that
static vacancies in a Heisenberg antiferromagnet enhance the staggered
magnetic moment within a few lattice spacings around the 
vacancy\cite{Bulut89}, while studies of models with only
short range antiferromagnetic (AF) correlations have shown
that correlations dynamically generate finite-range impurity
potentials from single site impurities\cite{scal}. 

In this paper we explore the role extended impurities have on transport
properties in an attempt to resolve the abovementioned discrepancies.
In particular, we re-examine the IR and the 
electronic Raman response of Bi-2212 including the effects of
electronic correlations on both inelastic and elastic scattering potentials,
and compare our results with the
extrapolated $T=0$ normal state resistivity and the crossover temperature
for $\lambda(T)$. We find that a consistent
picture emerges when we include the extended range
of impurity scattering as well as AF spin fluctuations.

Currently, knowledge of a $T-$ matrix formulation for disorder in correlated
systems is limited and generally approximate methods have been 
used\cite{ziegler}. Formally, one would need to include not only bare
impurity and interaction self energies responsible for elastic and inelastic
scattering, respectively, but also one must include terms in which mix
elastic and inelastic potentials. In this way, a purely static bare impurity
interaction changes in nature (can become dynamic) due to the inclusion of
many-body effects. Ziegler {\it et al.} in Ref. \cite{ziegler} discussed how
a point-like $s-$wave bare impurity potential can become extended due to the 
background of the correlated host. One can then proceed to calculate the 
$T-$ matrix by using a {\it renormalized}
Hamiltonian which describes the impurity potential in the correlated host.

Our starting point is the model Hamiltonian considered in Ref. 
\cite{anisimp}, which represents the effects of impurities
in a metal with short range AF order on a square lattice:
$$
H_{i}=\sum_{\{ {\bf l}\},\sigma,\pmb{$\delta$}} \left[{V_{0}\over{4}} 
n_{{\bf l},\sigma}+t_{I}(c_{{\bf l},\sigma}^{\dagger}
c_{{\bf l}+\pmb{$\delta$},\sigma}
+h.c.) + V_{1}n_{{\bf l}+\pmb{$\delta$},\sigma}\right].$$
The parameters $V_{0}$ and $V_{1}$ denote the on-site and extended 
impurity potentials, and $t_{I}$ denotes the effect of 
impurities on the electron hopping to the impurity site. 
Focusing on a two
parameter model using the specific relation $V_{1}=\alpha^{2}V_{0}/4$
and $t_{I}=\alpha V_{0}/4$, an analytic solution for the
single impurity $T$--matrix was obtained.
$\alpha$ is the control parameter which distinguishes between
point-like ($\alpha=0$) and extended $(\alpha \ne 0)$ 
impurity potentials. 

The algebraic solution for the impurity $T$--matrix 
is presented in Ref. \cite{anisimp}, where
the reader is referred to for details. In essence, the extended structure 
of the impurity potential 
requires a $4\times 4$ matrix formulation with respect to $s$, $p$, and $d$ 
scattering channels expressed in terms of the ${\bf k}$ dependent basis 
functions of the square lattice.
The impurity averaged self energy is determined via
$\hat\Sigma_{i}^{imp}({\bf k},i\omega)=
n_{i}\hat T_{\bf k,k}(i\omega)$, with $n_{i}$ the impurity
concentration. The self energy can be expanded in Pauli matrices 
$\hat\Sigma^{imp}=\sum_{i=0,3}\hat\tau_{i}\Sigma_{i}^{imp}$, with the
coefficients 
\begin{eqnarray}
&&\Sigma_{0}^{imp}
({\bf k},i\omega)={n_{i}V_{0}\over{\delta^{2}-\beta^{2}}}
\biggl\{-(1+\alpha\gamma_{\bf k})^{2}(d_{0}\beta+d_{3}\delta)+\label{eq2} \\
&& (\alpha\gamma_{\bf k}^{d})^{2}(s_{0}\beta-s_{3}\delta)\biggr\}
-{n_{i}V_{0}p_{3}\alpha^{2}\left([\gamma_{\bf k}^{p_{1}}]^{2}+
[\gamma_{\bf k}^{p_{2}}]^{2}\right)
\over{p_{0}^{2}-p_{3}^{2}+{a^{-}}^{2}}}, \nonumber \\
&&\Sigma_{1}^{imp}
({\bf k},i\omega)=\gamma_{\bf k}^{d}2\alpha n_{i}V_{0}
\biggl\{{(1+\alpha\gamma_{\bf k})a^{+}\delta\over{\delta^{2}-\beta^{2}}}
+{\alpha\gamma_{\bf k}a^{-}\over{p_{0}^{2}-p_{3}^{2}+{\bar a}^{2}}}
\biggr\}, \nonumber \\
&&\Sigma_{3}^{imp}
({\bf k},i\omega)={n_{i}V_{0}\over{\delta^{2}-\beta^{2}}}
\biggl\{(1+\alpha\gamma_{\bf k})^{2}(d_{0}\delta+d_{3}\beta)+
\nonumber \\
&&(\alpha\gamma_{\bf k}^{d})^{2}(s_{0}\delta-s_{3}\beta)\biggr\}
+{n_{i}V_{0}p_{0}\alpha^{2}\left([\gamma_{\bf k}^{p_{1}}]^{2}+
[\gamma_{\bf k}^{p_{2}}]^{2}\right)
\over{p_{0}^{2}-p_{3}^{2}+{a^{-}}^{2}}}, \nonumber
\end{eqnarray}
and $\Sigma_{2}^{imp}({\bf k},i\omega)=0$. 
Here $s_{0,3}, d_{0,3}, p_{0,3}, \delta, 
\beta$ and $a^{\pm}$ are constants\cite{constants} 
given by Eqs. (8-11) of Ref.
\cite{anisimp}. The ${\bf k}$-dependent basis functions for a
square lattice are 
$\gamma_{\bf k},\gamma_{\bf k}^{d}=[\cos(k_{x}a)\pm\cos(k_{y}a)]/2$, 
and $\gamma_{\bf k}^{p_{1}},\gamma_{\bf k}^{p_{2}}
=[\sin(k_{x}a)\pm\sin(k_{y}a)]/2.$ 
In the limit of $\alpha\rightarrow 0$,
the $T$-matrix becomes momentum independent and recovers the well
known results\cite{Hirsch88}. 

In the self consistent $T-$ matrix approximation, the non-interacting
Green's function is used
to calculate the self energy and then the new Green's
function is put back into the self energy calculation. 
The process is iterated until
convergence is realized, which occurs typically after only a few iterations.
We assume 
a strong on-site impurity interaction $V_{0}=8t$,
take $\epsilon_{\bf k}=-2t[\cos(k_{x}a)+\cos(k_{y}a)]+
4t^{\prime}\cos(k_{x}a)\cos(k_{y}a)-\mu$, 
$\Delta_{\bf k}(T)=\Delta_{0}(T)[\cos(k_{x}a)-\cos(k_{y}a)]/2$, a
weak coupling form for $\Delta_{0}(T)$ and choose 
$\Delta_{0}(T=0)=4T_{c}=0.4t$. Here and throughout,
lattice sizes of $32 \times 32$ up to 
$256 \times 256$ and typically
over 1000 frequencies were used. Our results showed little 
size effects above the $128 \times 128$ mesh.
The chemical potential $\mu$ was adjusted so that
the filling $\langle n \rangle= 0.825$. 

In the absence of vertex corrections,
the homogeneous Raman response and the real part of the
conductivity are
\begin{eqnarray}
&&\begin{array}{cc}
\chi^{\prime\prime}_{\gamma,\gamma}({\bf q}={\bf 0},\Omega) \\
\Omega\sigma^{\prime}_{xx}({\bf q}={\bf 0},\Omega)
\end{array}=
\sum_{\bf k}
\begin{array}{cc}
\gamma_{\bf k}^{2} \\
{j_{\bf k}^{x}}^{2}
\end{array}
\int {{\rm d}\omega\over{N\pi}}[f(\omega)-f(\omega+\Omega)]\nonumber \\
&&\times Tr\left\{
\begin{array}{cc}
\hat\tau_{3}\\
\hat\tau_{0}
\end{array}
\hat G^{\prime\prime}({\bf k},\omega)
\begin{array}{cc}
\hat\tau_{3}\\
\hat\tau_{0}
\end{array}
\hat G^{\prime\prime}({\bf k},\omega+\Omega)\right\},
\label{eq4}
\end{eqnarray}
where $f$ is the Fermi function and the current vertex $j_{\bf k}^{x}=
e \partial \epsilon_{\bf k}/\partial k_{x} =2tea \sin(k_{x}a)[1-2t^{\prime}
/t \cos(k_{y}a)]$.
The Raman response can be classified according
to elements of the $D^{4h}$ group:
\begin{eqnarray}
\gamma_{\bf k}(\omega_{I},\omega_{S})=
\cases{ b_{\omega_{I},\omega_{S}}[\cos(k_{x}a)-\cos(k_{y}a)]/4
, &$B_{1g},$\cr
 b_{\omega_{I},\omega_{S}}^{\prime}\sin(k_{x}a)\sin(k_{y}a)
, &$B_{2g},$\cr
 a_{\omega_{I},\omega_{S}}[\cos(k_{x}a)+\cos(k_{y}a)]/4
, &$A_{1g}.$\cr}
\label{eq3} 
\end{eqnarray}
If the light scattering is
non-resonant, the frequency dependence of the momentum
independent prefactors $b,b^{\prime}$ and $a$ can be safely neglected,
and we can adjust these prefactors to account for overall intensity.
It can be seen from the ${\bf k}-$dependence of the vertices
that the $B_{1g}$ response probes
qp dynamics around the Brillouin Zone (BZ) axes,
$B_{2g}$ probes the diagonals, and $A_{1g}$ is more of an
average around the BZ and involves pure density fluctuations and 
backflow\cite{footnote}.
Higher order terms of increasingly more anisotropic basis functions
could be considered but do not lead to appreciable differences
except for the $A_{1g}$ response (see Ref.\cite{bands} for details).
From here on we only consider the $B_{1g}$ and $B_{2g}$ channels.

\begin{figure}
\psfig{file=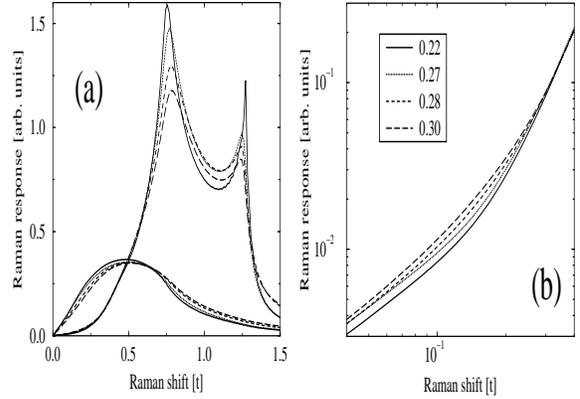,height=6.cm,width=8.cm,angle=0}
\caption[]{(a) $B_{1g}$ (upper set) and $B_{2g}$ (lower)
Raman response for $T=0.2T_{c}, \bar U=0$ and 1\% impurities. 
Here and in (b) the solid (dotted, dashed, long-dashed) line 
corresponds to $\alpha=0 (0.5, 0.75, 1)$. (b) Log-log plot of the
low frequency $B_{1g}$ Raman response determining
the crossover frequency $\omega^{*}$.
The legend gives the values of $\omega^{*}/t$.}
\label{fig1}
\end{figure} 

Results for the $B_{1g}$ and $B_{2g}$ response are plotted in Fig.
(\ref{fig1}a) for different values of $\alpha$.
As seen in our previous studies of the DOS\cite{anisimp}, 
turning on $\alpha$ even slightly leads to 
an effective increase in the strength of the impurity potential. The 
increased impurity resonance at the Fermi level
is manifest in the increasingly smeared spectra as well as the 
value of $\omega^{*}$, defined as the frequency 
where the cubic frequency dependence becomes sub-dominant\cite{ijmpb},  
as shown in Fig. (\ref{fig1}b). Since $\alpha=0.5$ corresponds to
a nearest neighbor impurity potential $V_{1}=V_{0}/16$, the presence of 
even a $6\%$ nearest neighbor potential leads to a $23\%$ increase in 
$\omega^{*}$. Therefore
a much smaller concentration of extended impurities is needed to have the same
effect as isotropic impurity scattering. The $B_{2g}$ channel is only slightly
affected by $\alpha$.

\begin{figure}
\psfig{file=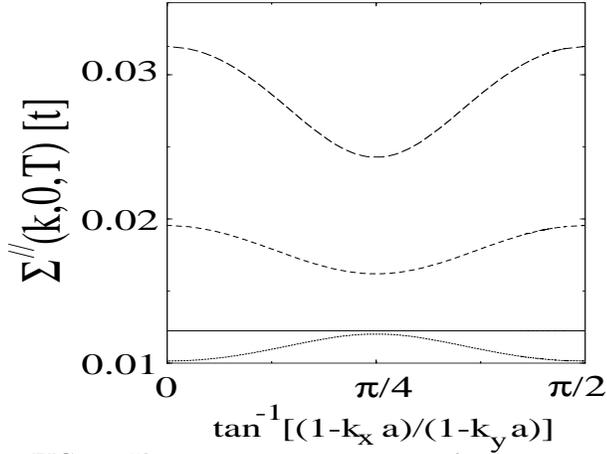,height=6.0cm,width=8.cm,angle=0}
\caption[]{The impurity scattering rate for momenta along the
Fermi surface
for $\alpha=0,0.24,0.48,$ and 0.72 (solid, dotted, dashed, long-dashed
lines, respectively).} 
\label{fig2}
\end{figure}

This is in accord with the impurity scattering being largest
near the BZ axes in our model and only minorly dependent on $\alpha$ near
the BZ diagonals. This is seen in Fig. (\ref{fig2}), which plots
the normal state $T=0$ impurity scattering rate around the Fermi surface
for different $\alpha$.
Increasing $\alpha$ from zero first decreases then increases the scattering
rate near the BZ axes while only mildly affecting the rate 
along the BZ diagonals.
Therefore we would expect the $B_{1g}$ channel to be more sensitive than
$B_{2g}$ to the growth of scattering for increased $\alpha$. 

\begin{figure}
\psfig{file=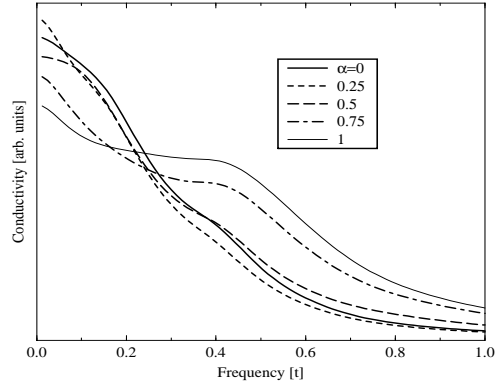,height=6.0cm,width=8.cm,angle=0}
\caption[]{The calculated IR at T=0.2T$_{c}$, 
$U=0$, and 2\% impurities for $\alpha=0,0.25,0.5,
0.75$ and 1 (solid, dashed, long-dashed, dash-dotted lines, and thin solid
lines,respectively).}
\label{fig3}
\end{figure}

As an important consequence, the IR
(which has similar weighting to $B_{2g}$) does not
pick up regions where the scattering changes rapidly
and is governed by small scattering along the BZ diagonals.
The IR in the superconducting state is shown in 
Fig. (\ref{fig3}) for different values of $\alpha$ at a fixed impurity
concentration. The IR does not appreciably change for
$\alpha$ up to 0.5. For larger $\alpha$ spectral weight is shifted away
from low frequencies as the
scattering rate increases over the entire BZ\cite{vertex}, and a stronger
shoulder near $\Delta_{max}$ develops.

We now consider inelastic scattering in order to address the 
intensity observed at high frequencies. The origin of the flat background
in IR and Raman at high frequencies has attracted a great deal of attention
involving both Fermi liquid and non-Fermi liquid based theories. A correct
approach would account not only for the featureless continuum but would be
able to describe the polarization dependence of the Raman cross section and
the differences between IR and Raman.
Raman has shown strong evidence of two-magnon features in the 
insulating as well as superconducting state, implying the role of spin
fluctuations as a source for inelastic scattering\cite{klein}. The
way in which spin fluctuations are included in calculating dynamic quantities
has also attracted a large amount of attention. The main problem has been
the degree in which strong local electron correlations are included and
represented. Effective Hamiltonian\cite{pines} 
approaches based on Bragg-like scattering
of quasiparticles for momentum ${\bf Q}=(\pi,\pi)$ have been successfully
employed to calculate various dynamic correlation functions, including
IR\cite{ps} and Raman\cite{NAFLtpd}. However, to a certain extent these
approaches have their limitation as they do not adequately capture the
strong multi-magnon scattering process required to approach the insulating
phase from the metallic side\cite{jrs}. 

Since here we are interested in only the low frequency behavior of the IR and
Raman response, the details of the dynamic scattering are not crucial in
that they only affect the response functions at larger frequencies. 
Therefore we take a simple route and include spin fluctuations in RPA,  
\begin{equation}
V({\bf q},i\Omega)={3\over{2}}
{{\bar U}^{2}\chi_{0}({\bf q},i\Omega)\over{1-\bar U
\chi_{0}({\bf q},i\Omega)}},
\label{eq5}
\end{equation}
where $\bar U$ is a phenomenological parameter [we choose $\bar U=2t$].
$\chi_{0}({\bf q},i\Omega)$ is the non-interacting spin susceptibility,
\begin{eqnarray}
&&\chi_{0}({\bf q},i\Omega)=\sum_{\bf k}
\biggl\{{a^{+}_{\bf k,k+q}\over{2N}} 
{f(E_{\bf k+q})-f(E_{\bf k})\over{i\Omega-(E_{\bf k+q}-E_{\bf k})}}
+ {a^{-}_{\bf k,k+q}\over{4N}}\\
\label{eq6}
&&\times
\left[{1-f(E_{\bf k+q})-f(E_{\bf k})\over{i\Omega+E_{\bf k+q}+E_{\bf k}}}
-{1-f(E_{\bf k+q})-f(E_{\bf k})\over{i\Omega-E_{\bf k+q}-E_{\bf k}}}\right]
\biggr\}.\nonumber
\end{eqnarray}
Here 
$E_{\bf k}^{2}=\epsilon_{\bf k}^{2}+\Delta_{\bf k}^{2}$ 
and the coherence factors are 
$a^{\pm}_{\bf k,k+q}=1\pm {\epsilon_{\bf k+q}\epsilon_{\bf k}+
\Delta_{\bf k}\Delta_{\bf k+q}\over{E_{\bf k+q}E_{\bf k}}}$. 
This yields a self energy
\begin{eqnarray}
&&\hat\Sigma^{\bar U}({\bf k},i\omega)=-\int{dx\over{\pi N}}
\sum_{\bf q} V^{\prime\prime}({\bf q},x){1\over{2E_{\bf k-q}}}
\nonumber\\
&&\biggl[{E_{\bf k-q}\hat\tau_{0}+\epsilon_{\bf k-q}\hat\tau_{3}+
\Delta_{\bf k-q}\hat\tau_{1}\over{E_{\bf k-q}+x-i\omega}}
[n(x)+f(-E_{\bf k-q})]- \nonumber \\
&&{-E_{\bf k-q}\hat\tau_{0}+\epsilon_{\bf k-q}\hat\tau_{3}+
\Delta_{\bf k-q}\hat\tau_{1}\over{-E_{\bf k-q}+x-i\omega}}
[n(x)+f(E_{\bf k-q})]\biggr].
\label{eq7}
\end{eqnarray}

\begin{figure}
\psfig{file=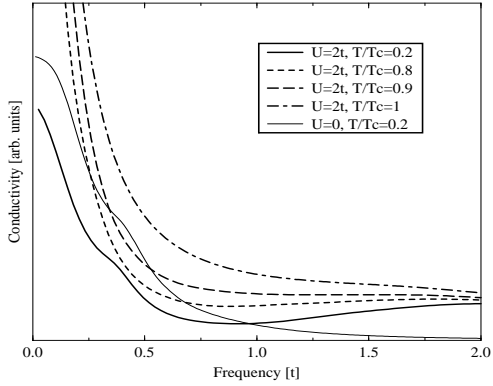,height=6.0cm,width=8.cm,angle=0}
\caption[]{The IR Conductivity for $T=0.2,0.8,0.9$, and $T_{c}$ (solid,
dashed, long-dashed, dash-dotted lines, respectively) for 2\% 
impurities and $\alpha=0.5$ and $U=2t$. The thin solid line is for 
$U=0, T=0.2T_{c}$ for comparison.}
\label{fig4}
\end{figure}

\begin{figure}
\psfig{file=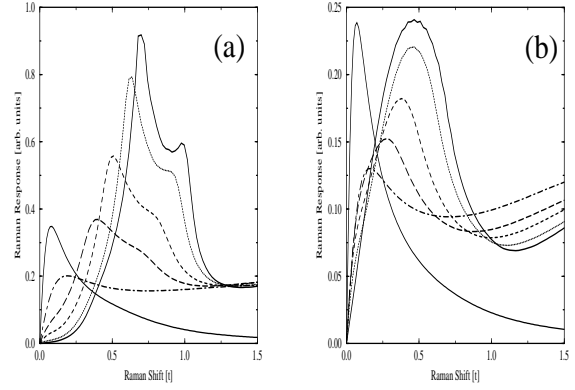,height=6.cm,width=8.cm,angle=0}
\caption[]{Raman response for the $B_{1g}$ (a) and $B_{2g}$ (b) channel for
$\bar U=2t, V_{0}=8t$, 2\% impurities, and $\alpha=0.5$ for $T/T_{c}=
0.2, 0.6, 0.8, 0.9, 1$ plotted 
as the thick solid, dotted, dashed, long-dashed,
and dotted-dashed lines. The thin solid line shows the
response at $T=T_{c}$ for $\bar U=0$ for comparison.}
\label{fig5}
\end{figure}

The convolution of momentum sums in Eq. \ref{eq7} is solved
numerically via Fast Fourier Transform, where we keep the full 
${\bf k}$-dependence and the real and imaginary parts of the self
energy. We found that neglecting the real parts of the self energy
and/or restricting momentum sums around the FS leads to a 
substantially smaller conductivity and misses a renormalization of
the conductivity peak to frequencies slightly away from $4\Delta$ in the
superconducting state\cite{Quinlan96}. 
Combining both $\hat\Sigma^{imp,\bar U}$,
the results for the IR and the
Raman response for the $B_{1g}$ and $B_{2g}$ channels 
are summarized in Figs. (\ref{fig4})
and (\ref{fig5}). For both quantities the spin fluctuations yield
a flat continuum at high frequencies in common with experiments
and various different theories which yield
a linear frequency dependence of the imaginary part of
the self energy. The temperature dependence there is minimal and all the
spectra converge to similar values by roughly $\Omega\sim 2t$. 
As the temperature is lowered, the low frequency IR falls in magnitude
and develops a shoulder at $\sim \Delta_{max}$ and a weak peak
at $\sim 4\Delta_{max}$, while the
spectral weight in the $B_{1g}$ and $B_{2g}$ channels reorganize
from low frequencies to higher frequencies at $\sim 2\Delta_{max}$ and
$0.65\Delta_{max}$, respectively. Strong peaks associated
with pair breaking and the van Hove structure ($\sim t$)
appear in the $B_{1g}$ channel and become less pronounced as
the temperature is increased due to the growth of spin fluctuation
scattering. We note that at present there is no experimental indication of
a peak in the $B_{1g}$ which could be associated with a van Hove feature
\cite{bi2201}. In our calculations further smearing of the van Hove peak 
is expected if dispersion is added in the $c-$direction or if stronger 
interactions are used which produce larger inelastic scattering 
at large frequencies. Moreover, a more correct multi-magnon approach would
also wash out structure at higher frequencies.

\begin{figure}
\psfig{file=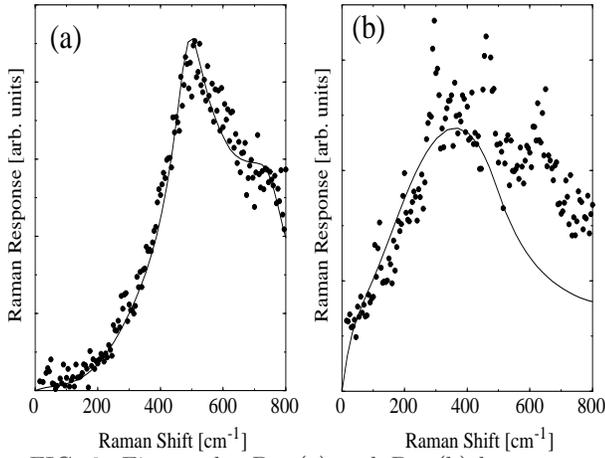,height=6.cm,width=8.cm,angle=0}
\caption[]{Fits to the $B_{1g}$ (a) and $B_{2g}$ (b) low temperature
spectra on Bi-2212 (T$_{c}$=86K) 
taken by R. Hackl {\it et al.} in \cite{bi2201}.}
\label{fig6}
\end{figure}

\begin{figure}
\psfig{file=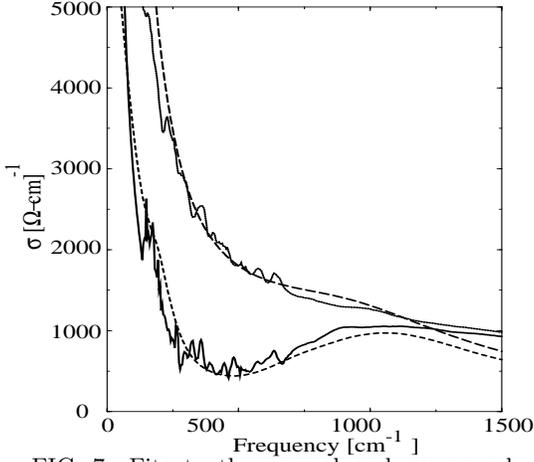,height=6.cm,width=7.cm,angle=0}
\caption[]{Fits to the normal and superconducting IR
on Bi-2212 (T$_{c}$=93K) 
taken by N. L. Wang {\it et al.}\cite{wang}.}
\label{fig7}
\end{figure}

Fits to the Raman spectra measured in the superconducting state at 
$T=0.5T_{c}$ of an as-grown sample of Bi-2212 are presented in 
Fig. \ref{fig6}, while fits to the IR in both the normal and 
superconducting state on a similar sample with a slightly higher $T_{c}$
by taken by N. L. Wang {\it et al.}\cite{wang} 
are shown in Fig. \ref{fig7}. 
Here we have taken the parameters used in Fig. \ref{fig5},
have adjusted the prefactors
$b/b^{\prime}=1.46$ to account for the relative Raman intensities.
For the IR, we use the $c-$axis lattice spacing of $30\AA$ for
Bi-2212 containing 2 CuO$_{2}$ bilayers to
convert the 2D IR to 3D. We find that theory underestimates the IR
scale by only a factor of 1.5 (the fits in Fig. \ref{fig7} are scaled
by this factor).
We used slightly different values of $t$ for Raman (81meV) than
IR (69 meV), but equally good fits are obtained if we used different
values for $\Delta$. The results
agree exceptionally well with the measured spectra especially
at low frequencies where
the effects of impurities are dominant. The agreement lessen to
to only a qualitative level for $\Omega > 1000$ cm$^{-1}$
due to the small
degree of spin fluctuations included (in RPA) and points to an inadequate
description of the normal state. 

Finally, we consider how the fitting 
parameters compare to transport data. Assuming a Drude model for
$\rho(T=0) (\approx  10 \mu\Omega$-cm) and a plasma
frequency of 1.2 eV\cite{forro} 
implies a scattering rate $1/\tau_{imp}= 15 $cm$^{-1}$, while the 
penetration depth
$T$ to $T^{2}$ crossover\cite{pene} 
measured in the same sample\cite{waldmann} as
the Raman data gives 12
cm$^{-1}$. Previous Raman fits using isotropic impurity
scattering only and a Fermi surface restricted approach 
required $1/\tau_{imp}=2\Gamma=72$ cm$^{-1}$\cite{ijmpb}. Our calculation
for $\alpha=0.5$ yields $1/\tau_{imp}^{ave}=23$ and 20 cm$^{-1}$ for the
FS averaged impurity scattering rate for the 
$t=81$ and 69 meV, respectively. Given the 
uncertainty in estimating $\rho(T=0)$ and $\omega_{pl}$, this in favorable
agreement with existing measurements.

In summary we have shown how the inclusion of electron correlations in
both inelastic and elastic scattering potentials lead to a consistent
description of the channel dependent 
Raman and IR response and lead to better fits than previously achieved.
The overall intensity of the IR can be accounted for and 
the effect of extended impurities on the low frequency
behavior of the Raman response can resolve the 
discrepancy between large impurity scattering rates needed previously for 
IR and Raman fits and the small rates needed for transport. 

\acknowledgements
Acknowledgment (T.P.D.) is made to the Donors of the Petroleum
Research Fund, administered by the American Chemical Society, for
partial support of this research. We have benefitted from discussions
with J. C. Irwin and R. Hackl, and we thank R. Hackl and
N. L. Wang for sharing the data presented in Figs. \ref{fig6} and \ref{fig7},
respectively.


\begin{references}
\bibitem{vertex}
Vertex corrections
must be included for finite $\alpha$. However, 
these corrections are proportional to $\alpha^{2}$ and thus
not important for small $\alpha$.
\end{references}
\end{document}